\documentclass[cits]{PoS}
\usepackage{graphicx, subfigure}
\usepackage{amsmath}
\usepackage{amsfonts}
\usepackage{amssymb}

\title{Traces of a fixed point:\\Unravelling the phase diagram at large Nf}

\ShortTitle{Traces of a fixed point}

\author{\speaker{Albert Deuzeman}\\
        Centre for Theoretical Physics\\
        University of Groningen \\
        9747 AG, Netherlands\\
        E-mail: \email{a.deuzeman@rug.nl}}
\author{Maria Paola Lombardo\\
        INFN -- Laboratori Nazionali di Frascati\\
	I-00044, Frascati (RM), Italy\\
        E-mail: \email{mariapaola.lombardo@lnf.infn.it}}
\author{Elisabetta Pallante\\
        Centre for Theoretical Physics\\
        University of Groningen\\
        9747 AG, Netherlands\\
        E-mail: \email{e.pallante@rug.nl}}

\abstract{With a sufficiently high number of fundamental
fermionic flavours present, Yang-Mills theory develops an infrared
fixed point and becomes (quasi-)conformal in nature. The range of
flavour numbers for which this occurs defines the conformal window,
the lower limit of which has yet to be determined. We studied the
phase diagram of SU(3) Yang-Mills theory with twelve flavours of
staggered fermions. Here we present evidence for the occurrence of a
bulk transition towards a chirally broken phase and the existence of a
Coulomb phase on the symmetric side of this transition, using results 
from the measurements of the chiral condensate and spectrum, leading to
the determination of a positive sign of the beta function. Assuming the
validity of the Appelquist-Miransky-Yamawaki scenario, this implies 
the existence of a conformal window that comprises the theory under
investigation.}

\FullConference{The XXVII International Symposium on Lattice Field Theory - LAT2009\\
		 July 26-31 2009\\
		 Peking University, Beijing, China}

\begin{document}
There has been a recent surge in interest in the physics of gauge theories with a relatively large number of fermionic degrees of freedom.
Research is ongoing on theories with different gauge groups and matter content in several representations \cite{Fleming:2008gy, Ryttov:2007sr}. 
The common factor of interest for all of these is the possible appearance of so called Banks-Zaks fixed points, first described in
two seminal papers of the late seventies and early eighties \cite{Caswell:1974gg, Banks:1981nn}. The central observation of these papers was 
the appearance of a second zero before the loss of asymptotic freedom in the two-loop beta function of an SU(3) gauge theory with $N_f$ massless 
fermions in the fundamental representation.  

Restricting ourselves to the latter case, the beta function to second order in perturbation theory, the highest order to which all coefficients 
are universal, is given by
\begin{eqnarray}
&&\beta (g) = \mu \frac{dg}{d\mu}= -b_0 g^3 - b_1 g^5 + O(g^7),\nonumber\\
&&b_0 = \frac{1}{16\pi^2}\left(11-\frac{2 N_f}{3}\right),\, b_1 = \frac{1}{(16\pi^2)^2}
\left(102-\frac{38 N_f}{3}\right).\nonumber
\end{eqnarray}
This equation has a root for $N_f^* > 8.05$ at $g^{*2}= -b_0/b_1$, causing, at this perturbative level, the appearance of an infrared fixed 
point up until $N_f^c = 16.5$ when asymptotice freedom is lost. Physically, a IR fixed point implies a phase in which chiral symmetry is never
spontaneously broken, as the coupling reaches an asymptotic value at infra-red scales and the theory turns quasi-conformal.

The actual value of the critical number of flavours $N_f^*$ is modified by the non-perturbative dynamics of chiral symmetry breaking and its 
restoration \cite{Miransky:1996pd}, necessitating an approach beyond perturbation theory. Analytical approximations have
been done, generally favoring a number larger than the perturbative prediction \cite{Appelquist:1999hr, Braun:2006jd, Ryttov:2007cx}, but with 
no consensus on the actual value. Lattice studies have concluded that $N_f=8$ lies within the hadronic phase of QCD \cite{Appelquist:2007hu, 
Appelquist:2009ty, Deuzeman:2008sc, Fodor:2009wk, Jin:2009mc}, but the case of $N_f=12$ is debated. A study based on the Schr\"odinger functional 
\cite{Appelquist:2007hu, Appelquist:2009ty} has concluded that $N_f=12$ should already be in the conformal window, but other groups have reported 
indications of a regular hadronic phase and spontaneous chiral symmetry breaking instead \cite{Fodor:2009wk, Jin:2009mc, Hasenfratz:2009kz}.
We here report on an investigation of the nature of SU(3) Yang-Mills theory with twelve flavours of quarks, employing the physics of phase 
transitions, to appear in print shortly~\cite{Deuzeman:2009mh}.

\section{\label{sec:setup}Setup and implementation}

\begin{figure}
\center
\includegraphics[width=10.0truecm]{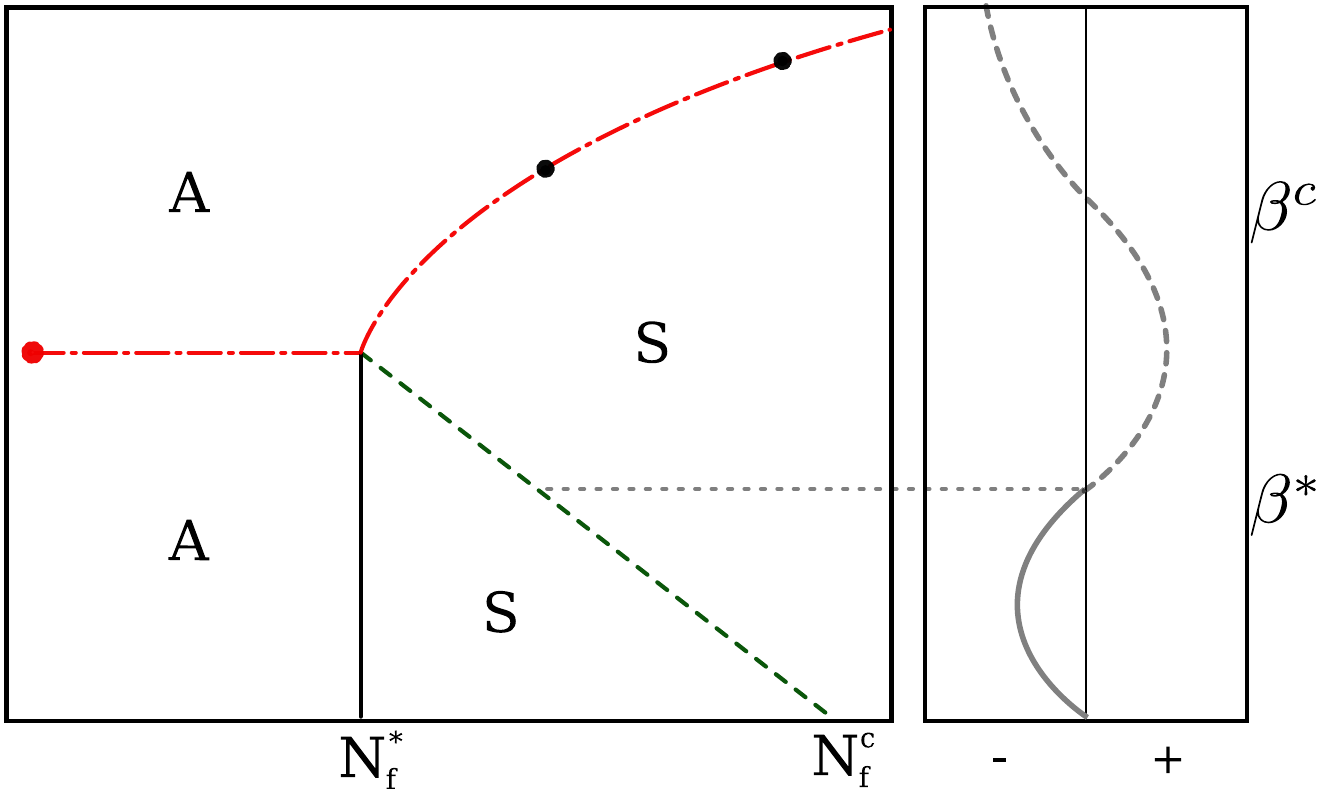}
\caption{\label{fig:manyphases} {
Diagram of the phases of an $SU(3)$ gauge theory encountered through varying the flavor content $N_f$ in the fundamental representation and the 
strength of the coupling $g$. Theories for $N_f < N^*_f$ are QCD-like, while for $N^c_f > N_f > N^*_f$  develop a conformal phase. S and A refer 
to chirally symmetric and asymmetric, respectively. The dot-dashed (red) line is the line of lattice bulk transitions. We observed a bulk 
transition at $N_f=12$ and also $16$ (dots). The dashed (green) line is the conformal critical line, qualitatively indicating the location of 
the IRFP's. The line at $N_f=N_f^*$ indicates a conformal phase transition with abrupt change of the spectrum, predicted in reference
\cite{Miransky:1996pd}. The non-perturbative beta function on the conformal side is also shown. }}
\end{figure}

In the setup of this study, we took guidance from the scenario painted in \cite{Miransky:1996pd}, a diagramatic depiction of which 
is given in figure~\ref{fig:manyphases}. For a given value of the flavour number $N_f > N^*_f$ and in the limit of massless quarks, decreasing 
$\beta$ will cause one to cross the critical value and go from the chirally symmetric (S) and asymptotically free phase to a phase that is still 
symmetric, but not asymptotically free. The green dashed line connects these roots of the beta function. The numerical values for which these 
occur are action dependent, but since the presence of such a fixed point is physically observable, it will appear for some value at any choice
of the action. Note, however, that the beta function is not universal away from fixed points with diverging correlation lengths and one can 
therefore not exclude the appearance of additional, spurious fixed points~\cite{Damgaard:1997ut}. At even smaller values of $\beta$, the lattice 
regularized theory could in general exhibit a transition to a strongly coupled chirally asymmetric (A) phase, driven purely by the value of 
coupling constant itself.

Outside the conformal window, in contrast, the theory will break chiral symmetry at zero temperature, but exhibit a thermal phase transition in 
the continuum to a high temperature chirally symmetric phase. Thus, as was argued in~\cite{Deuzeman:2008sc}, observing such a transition implies 
the absence of an infra-red fixed point. To discriminate between a true thermal transition of the hadronic phase and the aformentioned bulk 
transition of the quasi-conformal phase, we observed the scaling of the transition with the physical temperature, given by 
$T=\left(a\left(\beta\right)N_\tau\right)^{-1}$. The two types of transition are distinguished by their zero temperature, or infinite $N_\tau$ 
limit. If we obtain $\lim_{N_\tau \to \infty} \beta_T \to \infty$, we are in the hadronic phase, while $\lim_{N_\tau \to \infty} \beta_T  = \beta_c$  
implies a bulk transition for a finite value of $\beta_c$.

The bulk transition is a lattice artifact, one that specific forms of improvement may push down to arbitrarily low values of the lattice 
coupling and possibly have vanish altogether. Its potential appearance is relevant in that it would exclude its thermal counterpart. However, 
it is not inconceivable that one observes a transition between two different chirally broken phases, leaving room for a weak thermal transition
at some smaller value of the coupling. This scenario can be excluded primarily by checking directly for chiral symmetry restoration. In addition,
the presence of a Coulomb phase, signalled by a positive value of the beta function, is a necessary and sufficient condition for the presence of
a Banks-Zaks fixed point in the discretized system.

Simulations were performed with twelve dynamical flavours of fermions in the form of three unrooted staggered fermions. A tree level Symanzik
improved gauge action was used to suppress lattice artifacts. Fermions were Naik improved for an overall $\mathcal{O}(a)$ improvement that should
be valid in any phase of the theory. Allowing for the appearance of phases outside of regular hadronic QCD, other types of improvement may 
introduce unphysical terms in the continuum lattice action. Specifically, the Lepage-Mackenzie tadpole improvement scheme, an obvious choice when
attempting to observe thermal scaling for a phase transition, but derived from a perturbative expansion~\cite{Lepage:1992xa}, may introduce 
unphysical effects when used in the absence of asymptotic freedom. Taste breaking effects were estimated by comparing different pion sources and
found to be of the order of 40\%. By increasing pion masses, the effect of taste breaking contributions should be to partially freeze chiral
degrees of freedom and in this way effectively lower the number of fermionic degrees of freedom. Any conclusions drawn from the measurements
presented here will therefore specifically apply to twelve staggered degrees of freedom at finite mass, with a non-trivial continuum and chiral
limit. However, the shielding effect of the fermions responsible for any fixed point can only increase in those limits and if indications of its
existence are indeed found, they will carry through.

Runs were done on several volumes, with explicit checks on the size of finite volume effects for the high precision measurements needed to
perform the chiral extrapolation by comparing the results of three volumes and finding agreement within statistical errors for the largest two.
Final fits were based on $24^4$ volumes. Spectral measurements were done on $16^3\times24$ volumes, introducing a moderate amount of finite volume
corrections to the values obtained that tend to increase and equalize the obtained masses. Those effects, as long as they are not dominating, will
wash out the signal obtained for the sign of the beta function, but should not be expected to change it qualitatively.

\section{\label{sec:lattice}Results}

\begin{figure}
\center
\includegraphics[width=12.0truecm]{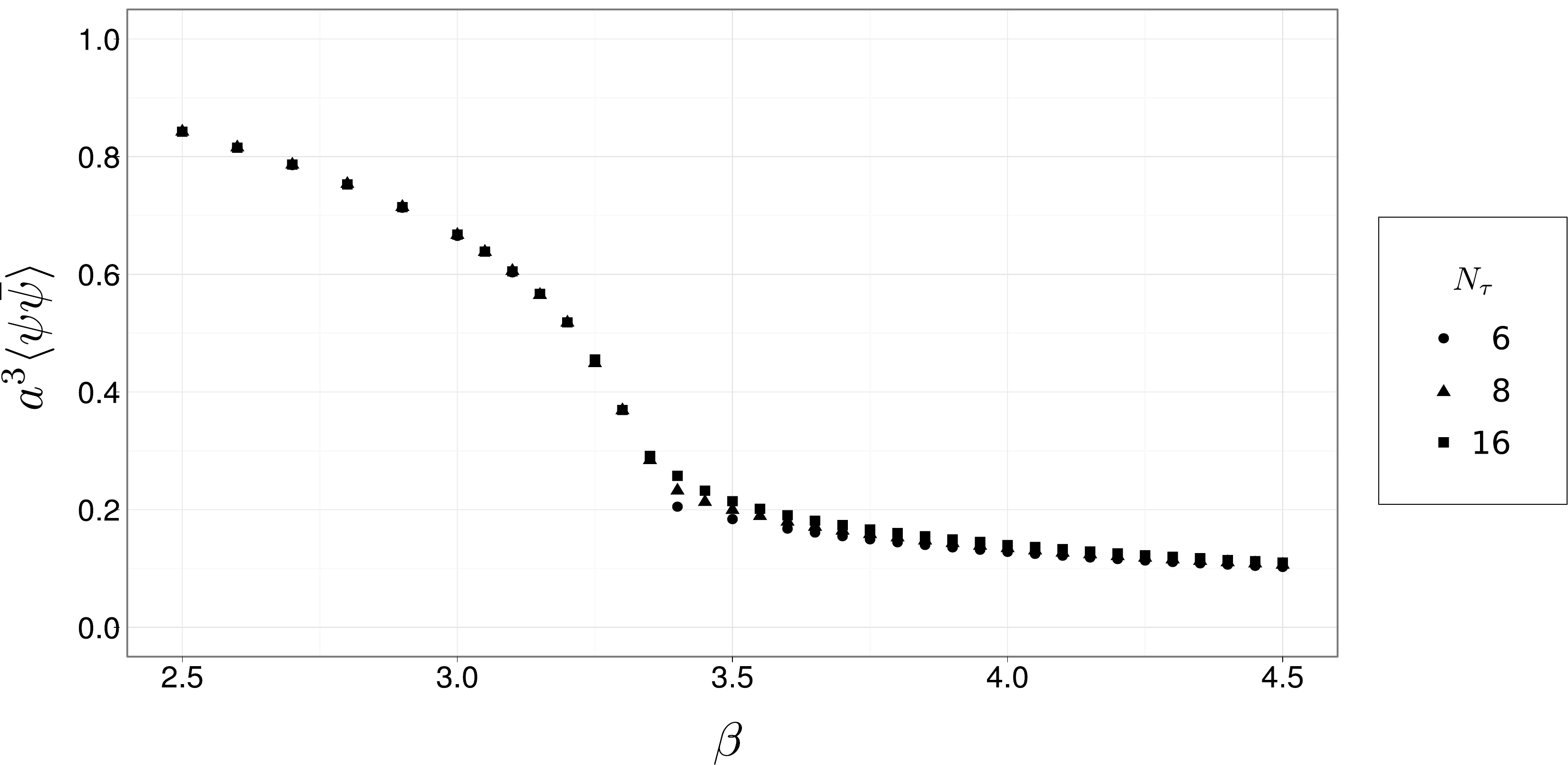}
\caption{\label{fig:bulk} {
The bulk transition in the chiral condensate for $am_q=0.05$ at several values of the temporal lattice extent $N_\tau$ . The location of the 
transition is identical, in spite of the physical temperatures differing by substantially. Simulation errors are within symbol sizes.
}}
\end{figure}

The size of the chiral condensate $\langle\bar\psi\psi\rangle$ was determined over a series of values of the lattice coupling constant $\beta$ 
(see fig.~\ref{fig:bulk}) and found to be a smoothly varying quantity for all measurements. In the range between $\beta=3.0$ and $\beta=3.5$, 
the condensate experienced a rapid acceleration in its decline, that tapered off again towards the region of weaker coupling. As observed above, 
such a transition can occur for actions within and outside of the conformal window, but is driven by a different quantity for either case. 
To see wether the physical temperature was the relevant factor, measurements were repeated for values of the lattice extent in the temporal 
direction $N_\tau$ differing by more than a factor of two. While the measurements on the weak coupling side of the transition show clear
thermal sensitivity, we observed no associated shift in the position of the transition itself. This implies that $\lim_{N_\tau \to \infty} \beta_T$
remains finite and the transition is of bulk nature.
\begin{figure}
\center
\includegraphics[width=12.0truecm]{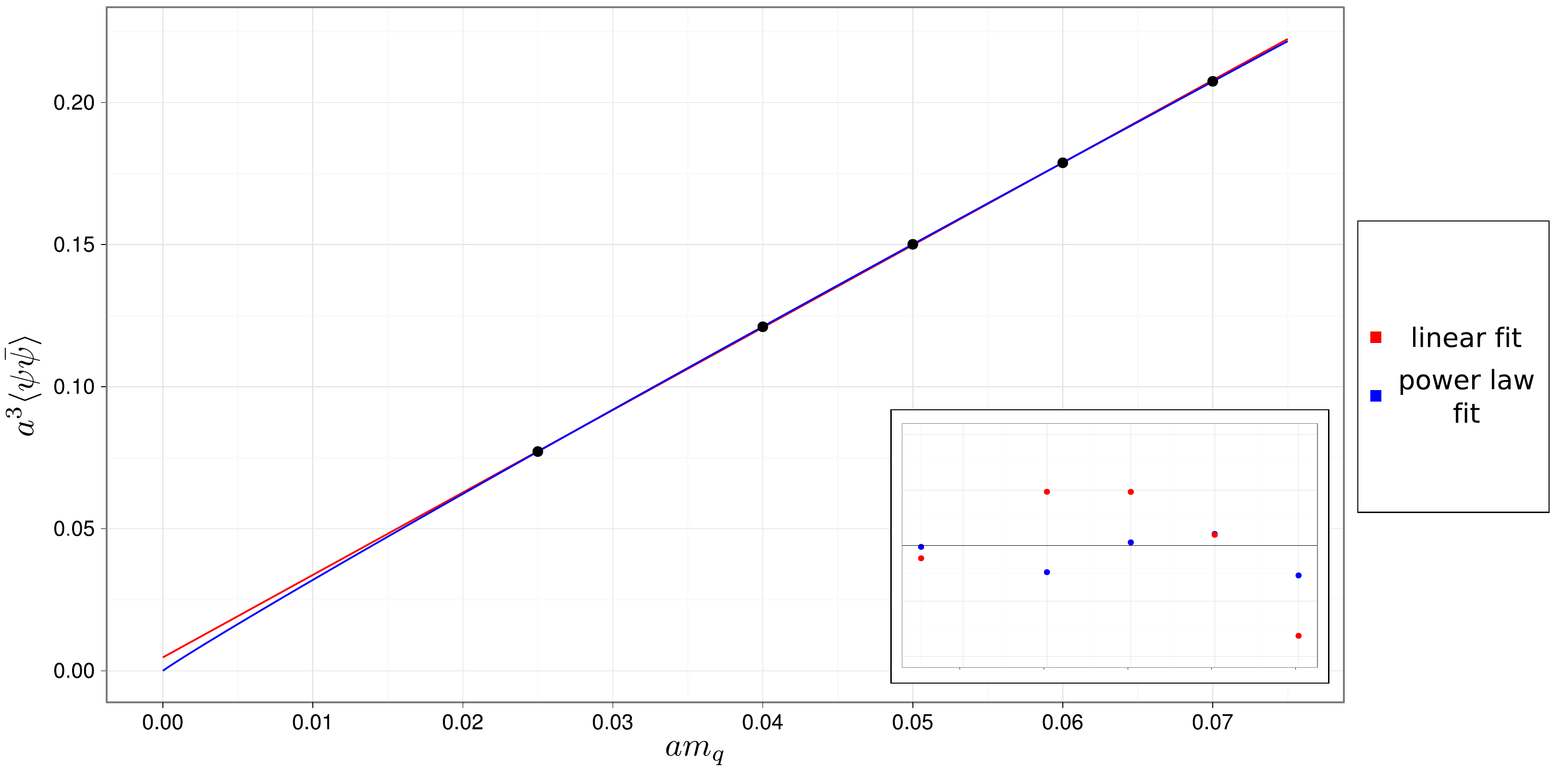}
\caption{\label{fig:condensate} {
Extrapolation to the chiral limit of the chiral condensate for $\beta=3.9$ with a linear formula (reduced $\chi^2=3.81$), giving
a small amount of residual chiral symmetry breaking, and a power law (reduced $\chi^2=0.86$), demonstrating a strong preference 
for the chirally symmetric scenario with a non-zero anomalous dimension.
}}
\end{figure}

To check for the restoration of chiral symmetry, an extrapolation to the chiral limit was performed. The dependence was approximately 
linear, but a slight systematic deviation of this behaviour could be observed. Most notably, higher masses for coupling showed clear upward 
deviations from linear behaviour close to the transition. This behaviour was reported earlier by Damgaard \textit{et~al.}~\cite{Damgaard:1997ut},
and is explained by both a shift in the position of the transition to weaker coupling and a broadening of the transition region as masses become
heavier, implying one observes remnants of the transition itself. Away from the transition these deviations rapidly weakened and the data developed 
a convex deviation from linearity. As these measurements were performed in an interacting system, one should generally allow for a non-zero 
anomalous dimension, as was, for example, shown for the chiral transition in the case of strongly coupling QED~\cite{Kocic:1990fq}. Additional 
high precision measurements were performed after the meeting, to be reported upon elsewhere, to perform a
quantitatively reliable chiral extrapolation. Figure~\ref{fig:condensate} displays an extrapolation to the chiral limit for a single value of 
the coupling of $\beta=3.9$ using, both with a linear form and allowing for an anomalous dimension. The difference between both forms is small
for the available masses, but the data prefer a scenario without spontaneous chiral symmetry breaking with a fitted exponent of 0.965(1).

Having observed a bulk transition to a phase that is compatible with the restoration of chiral symmetry, we tried to qualify the weakly coupling
phase further. One can attempt to distinguish a very weakly coupling hadronic phase from a Coulomb phase directly, by determining the sign of
the beta function. The beta function itself is a scheme dependent quantity, even when it is defined non-perturbatively, but the physical
nature of the fixed point should make the sign of the beta function a rather robust feature. We set up such a scheme by following reference~\cite{Damgaard:1997ut}, measuring the spectrum of light mesonic excitation at different 
values of the coupling constant and bare quark mass. Those coordinates in the two dimensional parameter space connected by renormalization group 
transformations should produce identical physical values, modulo a scaling factor. Subsequently, any dimensionless ratio of physical observables 
should be a constant, assuming that the system can be described by one parameter beta functions. Choosing the ratio of the ground state excitations 
in the pseudoscalar and vector channels arbitrarily and assuming a locally smooth behaviour of the function, the argument can be inverted to 
determine sets of parameters defining identical physics at different length scales.

The spread in values for the ratio is smaller than that observed for the case of $N_f=16$ in \cite{Damgaard:1997ut}, but the overall trend
is identical in that theories at weaker coupling and lower mass match to those at stronger coupling and higher mass. Because the signal-to-noise
ratio is lowered, the result is not as outspoken. One way of improving this is by interpolating the values of the ratio, effectively combining
statistics for the measurements. Such an extended analysis will be published~\cite{Deuzeman:2009mh}.
\begin{figure}
\center
\includegraphics[width=12.0truecm]{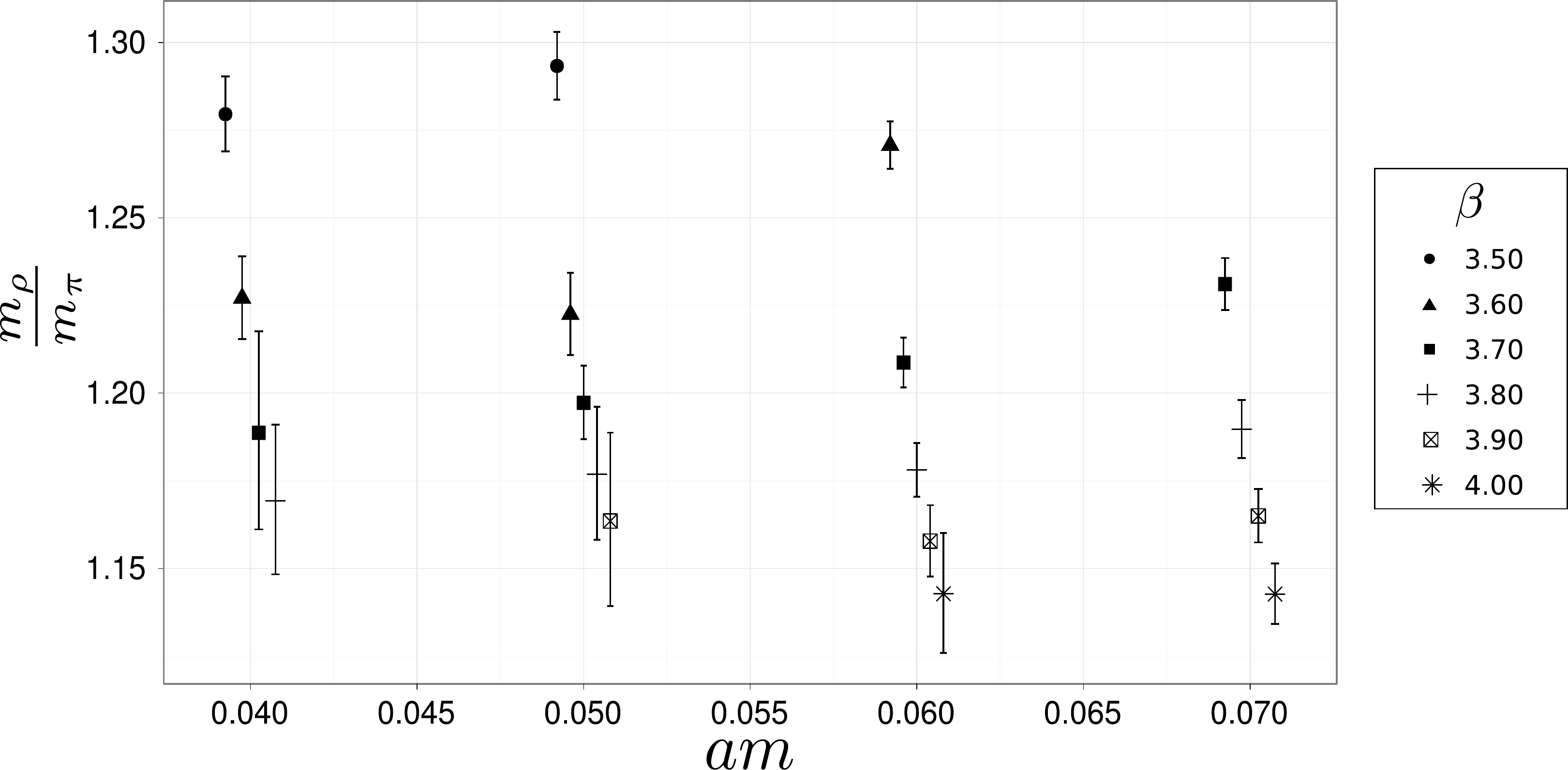}
\caption{\label{fig:Damgaard} The measured ratio of $m_\mathrm{vt}$ and $m_\mathrm{ps}$ as a function of the bare quark mass, for a series 
of values of the coupling constant $\beta$. Identical values of this ratio define theories extrapolating to the continuum limit, sampled at
different scales. The trend is towards higher values of $\beta$ with higher masses matching with lower values of $beta$ and lower masses,
indicating a decrease of the interaction strength for coarser lattices.}
\end{figure}
Comparing the ratios in \ref{fig:Damgaard}, the phase after the bulk transition is identified as a Coulomb phase. Assuming lattice artifacts 
are not dominating and under the assumption of continuity (which can be checked by scanning the parameter space towards zero coupling), 
this phase should connect smoothly to the perturbative function as the interaction weakens over larger distances. Since its sign is known 
to be negative perturbatively an intermediate zero has to be present.

\acknowledgments

This work was in part based on the MILC public lattice gauge theory code. We thank M. Bochicchio, S. Reker, F. Di Renzo, F. Sannino and C. deTar 
for comments and discussions. Computational resources were provided by the University of Groningen and the Dutch Stichting Nationale Computer 
Faciliteiten and are gratefully acknowledged.

\bibliographystyle{plain}

\end{document}